\documentstyle[11pt,paspconf,epsf]{article}

\begin{document}

\title{BIMA and Keck Imaging of the Radio Ring PKS~1830--211\altaffilmark{1}.}

\author{B. L. Frye}
\affil{Astronomy Department, University of California,
    Berkeley, CA 94720}

\author{F.    Courbin}
\affil{Institut   d'Astrophysique   et     de
G\'eophysique, Universit\'e  de  Li\`ege,\\
Avenue de  Cointe 5, B--4000 Li\`ege, Belgium.\\ 
URA 173 CNRS-DAEC,  Observatoire de Paris, F--92195
Meudon Principal C\'edex, France.}

\author{T. J. Broadhurst and W. J. W. Welch}
\affil{Astronomy Department, University of California,
    Berkeley, CA 94720}

\author{C.  Lidman}
\affil{European Southern Observatory, Casilla 19001, 
Santiago 19, Chile.}

\author{P. Magain\altaffilmark{2}}
\affil{Institut   d'Astrophysique   et     de
G\'eophysique, Universit\'e  de  Li\`ege,\\
Avenue de  Cointe 5, B--4000 Li\`ege, Belgium.}  

\author{M. Pahre and S.G. Djorgovski} 
\affil{Palomar Observatory, California Institute of
Technology, Pasadena, CA 91125, USA}



\altaffiltext{1}{Based on  observations obtained  at 
the W.M.  Keck Observatory, Hawaii, which is
operated  jointly  by the  California Institute  of Technology and the
University of California.} 
\altaffiltext{2}{Ma\^{\i}tre de Recherches au FNRS (Belgium)}




\begin{abstract}
We discuss BIMA (Berkeley Illinois Maryland Association) data and present new high quality optical and near-IR Keck
images of the bright radio ring PKS~1830--211.  Applying a powerful new
deconvolution algorithm we have been able to identify both images of the 
radio source. In addition we recover an extended source in
the optical, consistent with the expected location of the lensing galaxy.  The
source counterparts are very red, $I-K\sim7$ suggesting strong Galactic 
absorption with additional absorption 
by the lensing galaxy at $z=0.885$, and consistent with the detection of high
redshift molecules in the lens.

\end{abstract}




\section{Introduction}

   The bright radio source PKS~1830--211 (Jauncey et al.  1991) has attracted
much attention as the most detailed example of a lensed radio ring.  Of the
classically-lensed QSOs, its short time delay of 44 days (van Ommen et al.
1995) and clean lens geometry (e.g.  Subrahmanyan et al, 1990; hereafter S90),
make it a good candidate for measuring H$_{0}$.  The lens, a gas rich galaxy
at z=0.89, was discovered in the millimeter via molecular absorption (Wiklind
\& Combes 1996), which is seen towards only one of the two flat spectrum hot
spots (Wiklind \& Combes 1996, Frye et al. 1997).  A nearby saturated M-star
and heavy extinction along the line of sight (b=-5.7 degrees, Djorgovski, et
al.  1992; hereafter D92) has obscured the lens and the source from
identification.  In this paper we describe how the MCS deconvolution algorithm
(Magain, Courbin \& Sohy, 1997) was used to detect the 
counterparts of this bright radio ring lens in deep Keck optical and infrared
images.

\section{Deconvolution}

$I$ and $K$ band Keck data were obtained for this study.  The details of the
observations and reductions are presented in Courbin, et al.
1998.  The MCS deconvolution code (Magain, Courbin \& Sohy, 1997) was applied
to both the optical and IR images.  The limiting magnitudes 
of the images (3$\sigma_{sky}$
integrated over the whole object) are 24.0 in $I$ and 21.3 in $K$.

The deconvolution process is described in detail by Courbin, Lidman \&
Magain (1997) and was applied to the present data in an identical manner. As
output from the procedure, one gets a deconvolved image, decomposed into point
sources and a diffuse background, which is compatible with all the input images
included in the data set.  The photometry and the astrometry of the point
sources are also obtained as byproducts of the deconvolution (see Tables 1 and
2). Objects near the frame edges are not well-fitted.
(e.g. objects labelled 2,3 and 4 in Figure 1).  Note that object 1 is never
well-fit by a point source and leaves significant residuals after
deconvolution.  We therefore conclude that it is extended or that it is a very
strong blend of point sources.

\begin{table}[t]
\begin{center}
\caption{Summary of the photometry and astrometry the NE component and the SW
component of the lensed source (plus lensing galaxy), and photometry for the M-star. } 
\vspace*{10mm}
\begin{tabular}{c c c c}
\tableline
& M-star & NE Comp. & SW Comp.+Lens \\ \tableline
$I$ & $19.3 \pm 0.1$ & $22.0 \pm 0.2$ & $22.3 \pm 0.3$  \\
$K$ & $16.6 \pm 0.2$ & $15.1 \pm 0.1$ & $18.2 \pm 0.2$  \\ 
x($I$)&& $+0.08 \pm 0.01$ & $ +0.48 \pm 0.1$\\ 
y($I$)&& $-0.70 \pm 0.01$ & $ -1.15 \pm 0.1$\\ 
x($K$)&& $+0.06 \pm 0.01$ & $+0.59 \pm 0.05$\\ 
y($K$)&& $-0.54 \pm 0.01$ & $-1.20 \pm 0.05$\\ 
\tableline \tableline
\end{tabular}
\end{center}
\end{table}

\section{Results}

In Fig. 1 we present both the raw and deconvolved images in $I$ and $K$.  In
addition ESO $J$ and $K^{\prime}$ data were taken and are presented in Courbin,
et al. 1998.  The deconvolved images clearly show one red point source
at the position expected for the NE radio source of PKS~1830--211.  Another red
object is observed close to the position of the SW radio source of the lensed
system, but the extended nature of the source and the poor quality of
the PSF do not allow us to sort out its morphology.  The photometry and
astrometry of the field are presented in Tables 1 and 2 and Fig. 2.

The point source at the position of the NE radio source is likely to be 
the IR counterpart of the NE radio image of PKS~1830--211.  
With our high signal-to-noise we can show that its shape is compatible
with a point source, and its color, $I-K=6.9$, is much
redder than any ``normal'' star (e.g.  Koornneef, 1983).  
The I-band position is also within the
1$\sigma$ radio error bars. In Fig. 2 we align all positions with respect to
the M-star so that we can compare the different positions measured for the NE
and SW components at optical and IR wavelengths.

We spatially-resolved molecular absorption towards 
the two lensed images in the mm 
and measured the lensed source separation to be 0\farcs98 (see Fig. 3).  
In our optical and IR images the SW
component is $0.61\pm 0.13$\arcsec \ and $0.85\pm 0.09$\arcsec \ away from the
NE component respectively.  A plausible explanation for the apparent positional
shift between the optical, IR and radio positions is that
the SW component is a blend of two objects: the lensing galaxy and the heavily
reddened SW component seen in the radio images. 

 The flux ratio between the two lensed images of the source is 1 in the $I$ band
and $<$ 20 in $K$.
The combination of a reddened SW radio source plus blue lens can explain the large
flux ratios.  
Both NE and SW components are reddened,
making PKS~1830--211 another good example of a dusty lens, along with
MG~0414+0534 (e.g.  Annis \& Lupino, 1993) and MG1131+0456 (Larkin et al.
1994), the mean galactic extinction being far below the values considered for
PKS~1830--211. 


\section{Discussion}
The main result of our study is the detection of the optical and near-IR
counterpart to the NE radio source of PKS~1830-211 and the possible detection
of the SW component and lensing galaxy.  However, our
SW component candidate might be the lensing galaxy alone (its position is also
in good agreement with the predictions from the models calculated by S90 and
Nair, Narasimha \& Rao, 1993) or, given the crowding in the field, a red
galactic object almost coincident with the position of the SW radio source.  
The hypothesis of a demagnified third image of the source (S90)
between the 2 main lensed images is unlikely as 
in such a case extinction of the lens would have made it visible in the
IR.  Furthermore, the IR centroid of the SW component would have been shifted
towards ``object E'' rather than to the radio position of the SW component.

The higher contrast between the NE component and nearby M-star in the infrared
make near-IR spectroscopy necessary for finding the source z.  Deep, high
resolution near-IR imaging is needed to reveal the exact nature of the faint SW
component.  However, even at the highest resolution attainable, 0\farcs2-
0\farcs3 in the IR with HST (in particular in K where the SW component of
PKS1830-211 is best visible), deconvolution will be essential to discriminate
between the SW component candidate, the lensing galaxy and additional faint
galactic stars.


\acknowledgments

We thank Hy Spinrad for useful discussions
and Dick Plambeck for help with reducing the BIMA data.
The expert help of the staff at ESO and WMKO during the
observing runs was very much appreciated.
FC is supported by ARC 94/99-178 ``Action de Recherche Concert\'ee de la
Communaut\'e Fran\c{c}aise'' and P\^ole d'Attraction Interuniversitaire P4/05
(SSTC, Belgium).  SGD and MAP acknowledge support from the NSF PYI award
AST-9157412.


%
%

\newpage
\centerline{\bf FIGURE CAPTIONS}
\vskip 0.1in

Figure 1.  {Top row, from left to right:  1.  Field  of
4.0\arcsec\, around PKS~1830--211  observed with Keck~{\sc  ii} in the
I-band.   This image is   a stack of 6   frames  with a pixel  size of
0\farcs215 and  seeing of 0\farcs8.   
2.   Mean of 5
K-band images obtained with Keck~{\sc i} and  NIRC.  The pixel size is
0\farcs157 and the seeing is 0\farcs7.  
Bottom row, from left to right:  
1.  Simultaneous deconvolution of  the  6  I-band frames:
resolution   of  0\farcs215   and pixel size     of  0\farcs1075.  
2.  Deconvolution  of the
mean  of 5 NIRC  images:  resolution of 0\farcs157   and pixel size of
0\farcs0785.   In all the images North  is  up, East left. The M-star,
the NE QSO component candidate and the blended  SW component + lensing
galaxy  candidate  are  detected. The  scale   is given in  arcseconds
relative to the M-star.  The NE QSO is directly south of the M-star.}
\medskip

Figure 2.  {Positions observed for  the different objects detected in the
optical  and near-IR along with their 1$\sigma$ error bars,  
relative to  the M star.   The
large open  circles show  the geometry of  the   system in the  radio.
Their radius corresponds  to the error bars  quoted by S90.  The black
dots are from  the near-IR images  while the open circles indicate the
result obtained from the I-band data.}
\medskip

Figure 3.  {The 3 mm continuum map (center) showing the known double lensed
structure is well resolved with a separation of 0\farcs98 and a flux ratio of
$1.14\pm 0.05$.  Contour levels are spaced by 0.1 Jy/beam with the lowest value
at 0.1 Jy/beam.  The synthesized beamwidth is 0\farcs68 $\times$ 0\farcs45
(lower left), where the asymmetry is due to the low elevation of the source as
seen from the northern latitude of BIMA.  The spectral range covers the
redshifted $HCN$(2-1) and $HCO^+$(2-1) molecular transitions at 5 km/s
resolution, which are shown (inset) for both images. The molecular absorption
is detected only in the SW component and does not reach the base of the
continuum. Since the two images are similarly bright, the lack of absorption in
the NE spectrum confirms that the spill over between the two images is
negligible.}

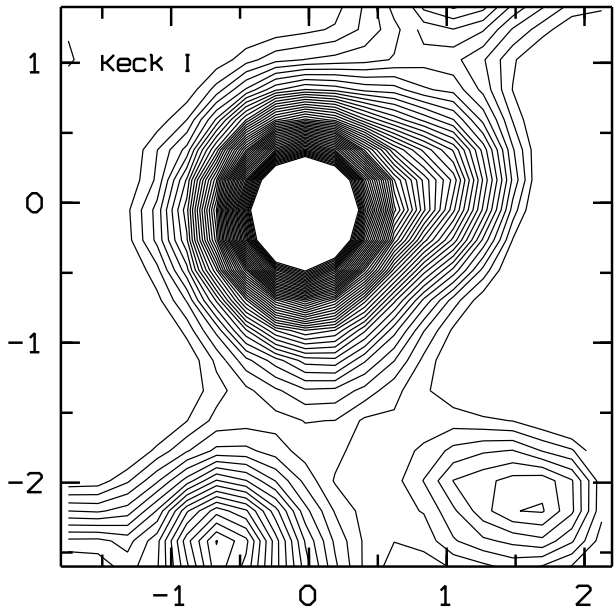
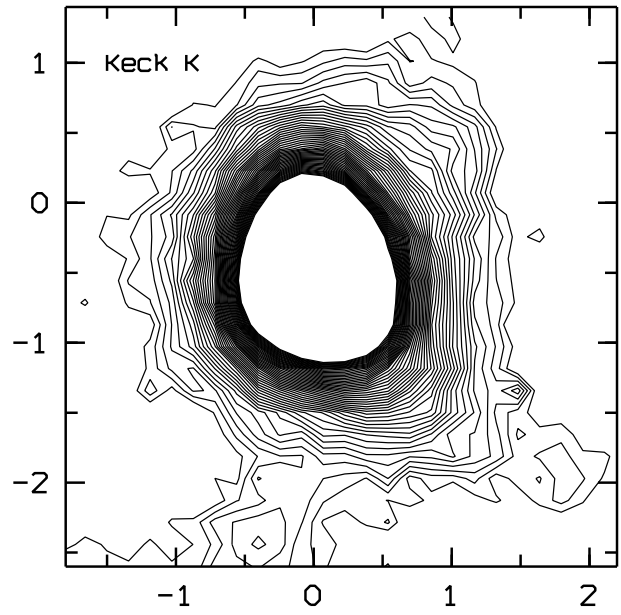
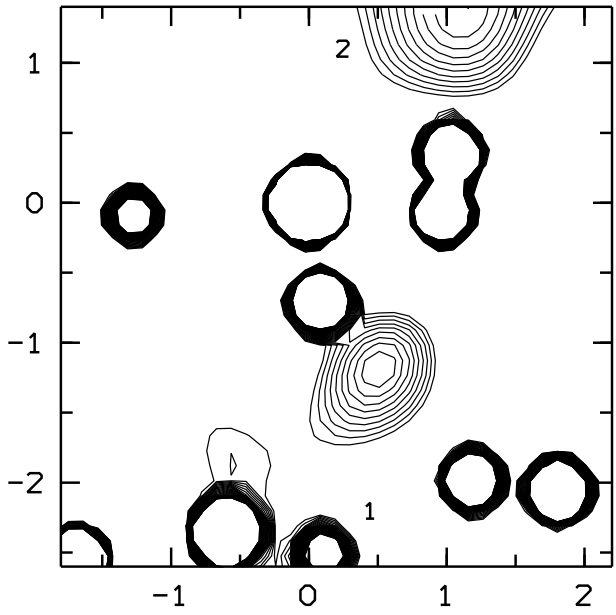
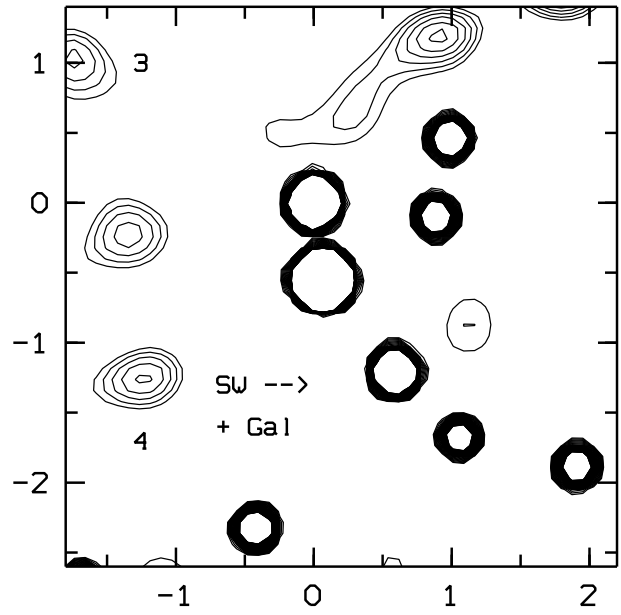
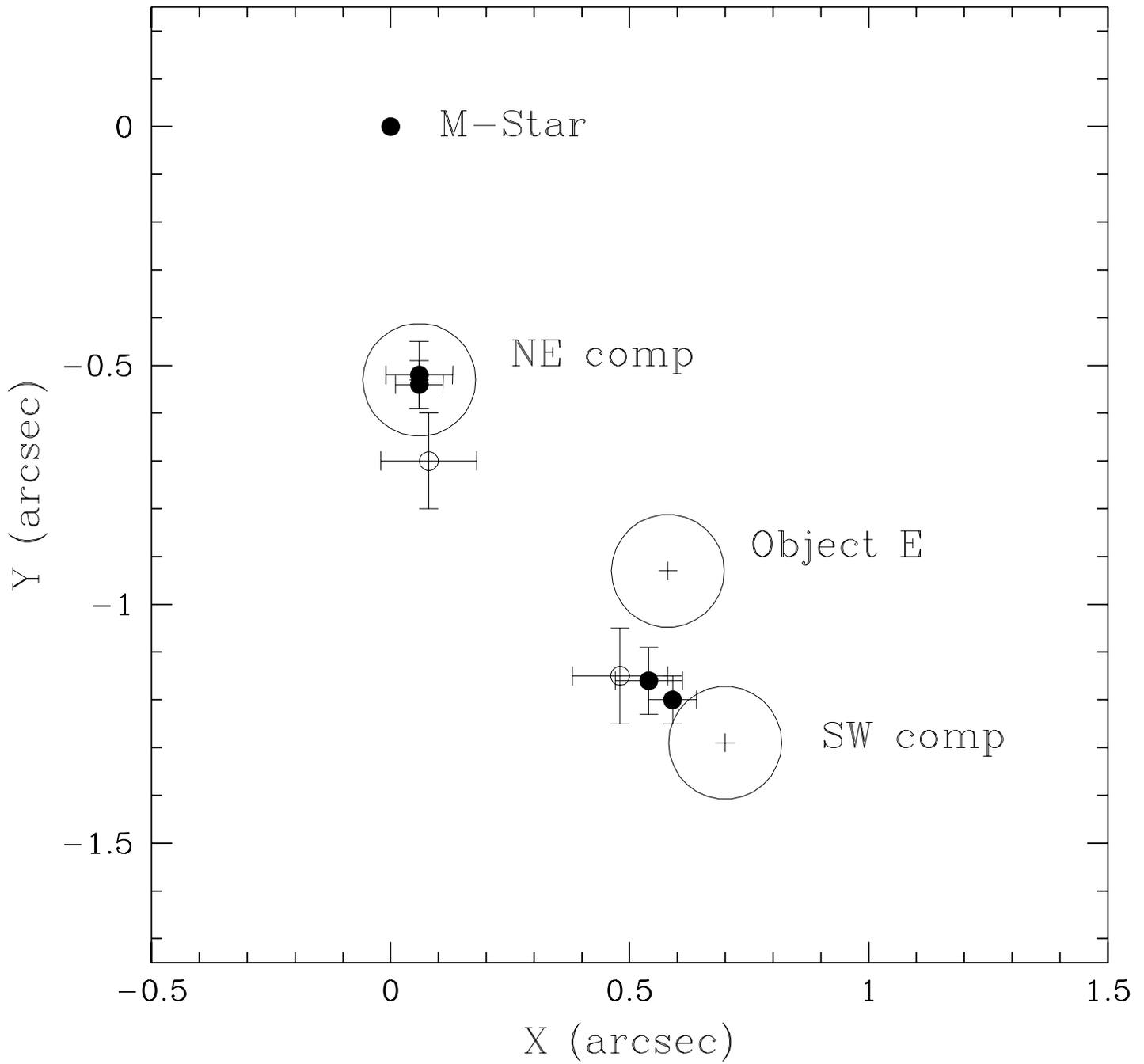
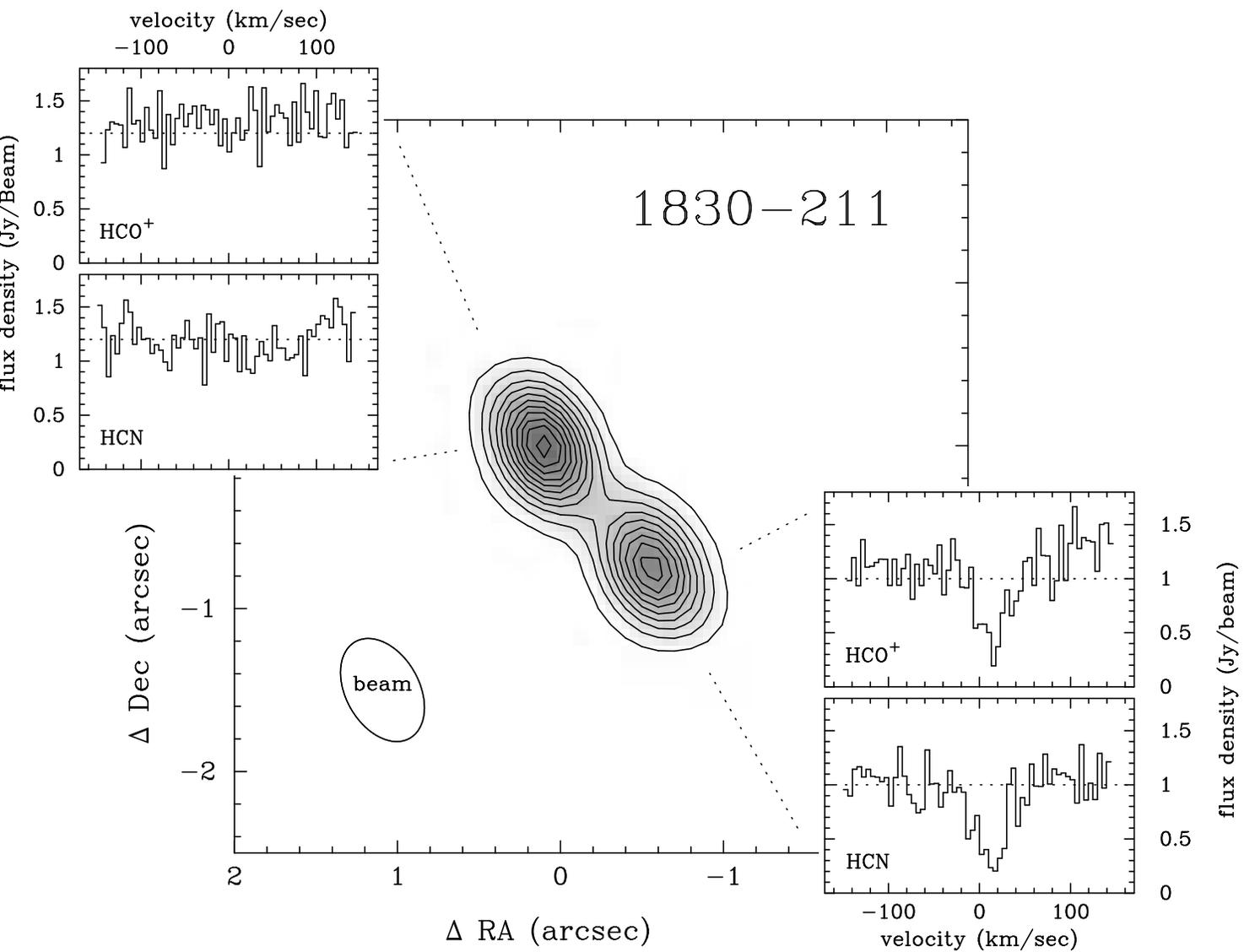
\end{document}